\documentclass[11pt]{article}
\usepackage[round,colon]{natbib}

\usepackage{graphics,epsfig}

\begin{document}

\title{
Plankton blooms in vortices: The role of biological and
hydrodynamic timescales. }

\author{ Mathias Sandulescu$^1$,
Crist\'obal L{\'o}pez$^2$, \\
Emilio Hern{\'a}ndez-Garc{\'\i}a$^2$,
Ulrike Feudel$^1$ \\ \
\\ \ \\
$^1$ Institute for Chemistry and Biology of the Marine Environment, \\
Carl-von-Ossietzky Universit\"at Oldenburg \\
D-26111 Oldenburg, Germany \\
$^2$ Instituto de F{\'\i}sica Interdisciplinar y Sistemas
Complejos-IFISC \\
(CSIC - Universitat de les Illes Balears) \\
E-07122 Palma de Mallorca, Spain \\
\ \\ \ \\
Published in\\
Nonlinear Processes in Geophysics {\bf 14}, 443-454 (2007)
\\http://www.nonlin-processes-geophys.net/14/443/2007/
\\under a Creative Commons License
}


\maketitle
\begin{abstract}
We study the interplay of hydrodynamic mesoscale structures and
the growth of plankton in the wake of an island, and its
interaction with a coastal upwelling. Our focus is on a mechanism
for the emergence of localized plankton blooms in vortices. Using
a coupled system of a kinematic flow mimicking the mesoscale
structures behind the island and a simple three component model
for the marine ecosystem, we show that the long residence times of
nutrients and plankton in the vicinity of the island and the
confinement of plankton within vortices are key factors for the
appearance of localized plankton blooms.
\end{abstract}

\section{Introduction}
 The interplay between hydrodynamic motion and
the distribution of marine organisms like phytoplankton and
zooplankton is a major challenge recently addressed  in numerous
studies \citep{Mann1991, Denman1995, Abraham1998, Peters2000,
Karolyi2000, Lopez2001b,
Lopez2001,Martin2002,Martin2003,Sandulescu2007}.

The growth of phytoplankton in the world's oceans depends strongly
on the availability of nutrients. Thus, one of the essential
factors controlling the primary production is the vertical
transport of nutrients. Coastal upwelling is one of the most
important mechanisms of this type. It usually occurs when
wind-driven currents, in combination with the Coriolis force,
produces Ekman transport, by which surface waters are driven away
from the coast and are replaced by nutrient-rich deep waters. Due
to this nutrient enrichment, primary production in these areas is
strongly boosted, giving rise also to an increase of zooplankton
and fish populations.

On the other side, the interplay between plankton dynamics and
horizontal transport, mixing and stirring has been investigated in
several studies recently \citep{Abraham1998, Lopez2001,
Hernandez2002, Hernandez2003b, Martin2003}. Horizontal stirring by
mesoscale structures like vortices and jets redistributes plankton
and nutrients and may enhance primary production
\citep{Martin2002,Hernandez2004}.
 Horizontal transport can
also initiate phytoplankton blooms and affects competition and
coexistence of different plankton species
\citep{Karolyi2000,Bracco2000}.

Vertical upwelling in connection with strong mesoscale activity
occurs in several places on Earth. One of these regions is the
Atlantic ocean area close to the northwestern African coast, near
the Canary archipelago. The main water current in this area flows
from the Northeast towards the Canary islands, in which wake
strong mesoscale hydrodynamic activity is
observed~\citep{Aristegui1997}. The interaction between the
vortices emerging in the wake of the Canary islands and the Ekman
flow seems to be essential for the observed enhancement of
biological production in the open southern Atlantic ocean close to
the Canary islands~\citep{Aristegui2004}. The aim of this paper is
to study the interplay between the redistribution of plankton by
the vortices and the primary production. In particular we focus on
the role of residence times of plankton particles in the wake of
the island. Though we believe that our study is relevant for
different areas in the world, we focus on the situation around the
Canary archipelago to be specific.

In this work we consider the coupling of the kinematic flow
introduced  in \citep{Sandulescu2006} to a simplified model of
plankton dynamics with three trophic levels, and study the impact
of the underlying hydrodynamic activity and the upwelling of
nutrients on primary production in different areas of the wake. In
this setup vortices have been reported to play an essential role
in the enhancement of primary production \citep{Sandulescu2007}.
Our main objective here is to analyze this mechanism in detail and
show that the extended residence times of plankton within vortices
are responsible for the observation of localized algal blooms in
them.

The organization of the paper is as follows. In section
\ref{general} we present the general framework of our system,
indicating the hydrodynamical and the biological model, as well as
their coupling. Our main analysis is devoted to the mechanism of
the appearance of a localized plankton bloom within a vortex (Sec.
\ref{results}). We study the residence times of plankton within
vortices and in the neighborhood of the island. Additionally we
clarify the role of the chaotic saddle embedded in the flow in the
wake of the island. Finally we summarize and discuss our results
in Sec. \ref{conclusions}.

\section{General framework: velocity field, plankton model and boundary
conditions}
\label{general}

\begin{figure} [th]
\begin{center}{
\includegraphics [angle=0, width=\textwidth]{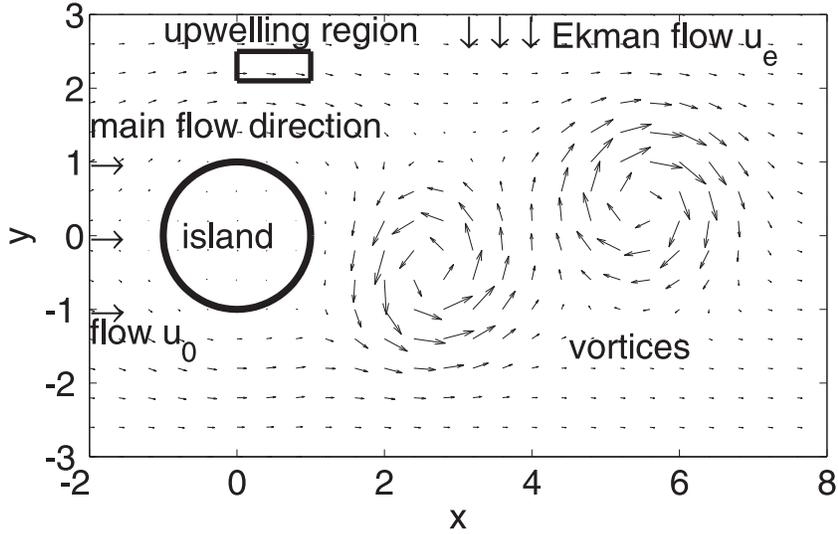}
}
\end{center}
\caption{The simplified island wake model setup.
\label{fig:area}}
\end{figure}

Our system consists of a hydrodynamic flow with an embedded
obstacle and vortices in its wake. The model contains also a
current perpendicular to the main flow that models an Ekman flow
coming from the coast, and  a nutrient-rich region at a distance
from the obstacle simulating a coastal upwelling zone. A sketch of
the model is shown in Fig. \ref{fig:area}. With this simplified
geometry we mimic the essential features of the hydrodynamic flow
in the Canaries (note that the whole Canary archipelago is
approximated by one cylindrical island). In particular, in the
wake of the obstacle strong mesoscale activity is observed in the
form of a periodic detachment of vortices, which then travel in
the main flow direction.

We use the kinematic model first developed by \citet{Jung1993},
which we modified by the introduction of the Ekman flow
\citep{Sandulescu2006}. This  model is coupled to a simple
population dynamics which features the interaction of nutrients
$N$, phytoplankton $P$ and zooplankton $Z$. The next two
subsections are devoted to the introduction of the hydrodynamic as
well as the biological model before discussing the results of
coupling both models to study the feedback between hydrodynamics
and phytoplankton growth.

\subsection{The hydrodynamic model}
\label{The velocity field}

We now introduce the velocity field. Details can be found in
\citep{Sandulescu2006}. The setup of our hydrodynamic model is
based on a horizontal flow pattern.  As Fig. \ref{fig:area} shows
the main current runs from left to right along the horizontal $x$
direction. The  center of the cylinder is placed at the origin of
the coordinate system. We consider a two-dimensional velocity
field which can be computed analytically from a stream function
$\Psi$. The velocity components in $x$- and $y$-direction and the
equations of motion of fluid elements are:
\begin{eqnarray}
 \dot x &=& v_{x}(x,y,t) = \frac{\partial}{\partial y} \Psi(x,y,t),\nonumber \\
 \dot y &=& v_{y}(x,y,t) = -\frac{\partial}{\partial x} \Psi(x,y,t).
\label{Strfcomp}
\end{eqnarray}

The stream function is  given by the product of two terms
\citep{Jung1993}:

\begin{equation}\label{StrfFG}
\Psi(x,y,t) = f(x,y) g(x,y,t).
\end{equation}

The first factor $f(x,y)$ ensures the correct boundary conditions
at the cylinder, $f(x,y)= 1-\exp\left[-a
\left(\sqrt{x^{2}+y^{2}}-1 \right)^2\right]$.
 The second
factor $g(x,y,t)$ models the background flow, the vortices in the
wake, and the Ekman flow $g ( x,y )=  -w h_{1}(t)g_{1}(x,y,t) +w
h_{2}(t)g_{2}(x,y,t) +u_0 s(x,y) y + u_E \Theta (x-1) x$. The
vortices  in the wake are of opposite sign but their maximal
vortex strengths are equal and denoted by $w$, and its shape is
described by the functions $g_i$ (see details in
\cite{Sandulescu2006}).
The characteristic linear size of the vortices is given by
$\kappa_0^{-1/2}$
 and  the characteristic ratio between the
elongation of the vortices in the $x$ and $y$ direction is given
by $\alpha$. The vortex centers move along the $x$ direction
according to $x_{1}(t) = 1+L\left( t/T_{c}\ {\rm mod}\ 1 \right)$
and $x_{2}(t) = x_{1}(t-T_{c}/2)$, and at values of $y_i(t)$
described below.
 Each vortex travels along the $x$ direction
for a time $T_{c}$ and disappears. The background flow moves in
the positive horizontal direction with a speed $u_0$. The factor
$s(x,y)$
describes the shielding of the background flow by the cylinder in
a phenomenological manner, using the same elongation factor
$\alpha$ as for the vortices. The Ekman drift, which is intended
to model the flow from the coast towards the ocean interior, is
introduced by considering an additional velocity of constant
strength $u_E$ in the $y$ direction acting only at $x$ coordinates
larger than $1$, i.e. just behind the island. This corresponds to
a stream crossing the vortex street towards negative $y$ values
beyond the cylinder.

Real oceanic flows are never perfectly periodic. Therefore we use
a non-periodic version of the kinematic flow just presented.
Non-periodicity is achieved by adding some randomness to the
vortex trajectories. Instead of moving along straight horizontal
lines, $y_1(t)=y_0$, $y_2(t)=-y_0$ ($y_0$ constant), the vertical
coordinates of the vortices move according to $y_1(t)=y_0+\epsilon
\xi(t)$, and $y_2(t)=-y_1(t)$, where, at each time, $\xi(t)$ is a
uniform random number in the range $[-1,1]$, and $\epsilon$ is the
noise strength.

The parameters of the model are chosen in such a way that they
represent properly the geophysical features of the Canary zone.
These values are given in Table \ref{hydroparameters}. To make the
model dimensionless we measure all lengths in units of the island
radius $r=25$ km and all times in units of the period $T_c=30$
days. For a discussion of all parameters we refer to
\citep{Sandulescu2006}, where the adaptation of the model to the
situation around the Canary islands is discussed in detail.

\begin{table}[ht] \label{hydroparameters}
\caption{List of parameters used in the hydrodynamical model}
 \begin{tabular}[t]{l l l}
 \hline
 parameter & value & dimensionless value\\
 \hline
 $r$ & 25 km & 1\\
 $u_0$ & 0.18 m/s & 18.66\\
 $\kappa_{0}^{-1/2}$ & 25 km & 1\\
 $\alpha$ & 1 & 1\\
 $w$ & $\approx 55 \times 10^3\ m^2/s$ & 200\\
 $T_c$ & 30 days & 1\\  $L$ & 6r~=~150 km & 6\\
 $a^{-1/2}$ & 25 km & 1\\
 $u_E$ & 0.02 m/s & 2\\
 $y_0$ & $r/2$=12.5 km & 0.5\\
 $\epsilon$ & 6.25 km  & 0.5\\
 \hline
\end{tabular}
\end{table}


\subsection{The biological model}

One can find in the literature a large variety of different models
used to analyse the dynamics of marine ecosystems. Their
complexity ranges from simple ones with only a few interacting
components \citep{Steele1981,Steele1992} to large comprehensive
ones \citep{Baretta1997}. We use a system which is based on a
three component model developed by \citet{Steele1992} and later
modified by \citep{Edwards1996} and \citep{Oschlies1999}.

The model describes the interaction of three species in a trophic
chain, namely nutrients $N$, phytoplankton $P$ and zooplankton
$Z$, whose concentrations evolve in time according to the
following dynamics:

\begin{eqnarray}
\frac{dN}{dt} =~F_N~&=& \Phi_{N} - \beta \frac{N}{k_{N}+N}P+
\nonumber \\&+& \mu_{N} \left((1-\gamma) \frac{\alpha \eta
P^{2}}{\alpha + \eta P^{2}}Z + \mu_{P}P + \mu_{Z}Z^{2}\right)
\nonumber
\\
\frac{dP}{dt} =~F_P~&=& \beta \frac{N}{k_{N}+N}P - \frac{\alpha
\eta P^{2}}{\alpha+ \eta P^{2}}Z - \mu_{P}P \nonumber
\\
\frac{dZ}{dt} =~F_Z~&=& \gamma \frac{\alpha \eta P^{2}}{\alpha +
\eta P^{2}}Z - \mu_{Z}Z^{2}
\label{BioDLG}
\end{eqnarray}

Let us now briefly discuss the meaning of the different terms (cf.
\citet{Oschlies1999} and \citet{Pasquero2004} for details): the
dynamics of the nutrients is determined by nutrient supply due to
vertical mixing, recycling by bacteria and consumption by
phytoplankton. Vertical mixing which brings nutrients from lower
layers of the ocean into the mixed layer is parameterized in the
biological model, since the hydrodynamical part considers only
horizontal transport of nutrients. For the vertical mixing we
assume  $N_0$ as a constant nutrient concentration below the mixed
layer. Thus the mixing term reads:

\begin{equation}\label{S}
\Phi_{N}=S(x,y)(N_0-N),
\end{equation}
where the function $S$ determines the strength of the upwelling
and will be discussed in more detail below. The nutrients are
consumed by phytoplankton with a saturation characteristics
described by a Holling type II functional response. The recycling
by bacteria is modelled by the last three terms in the bracket. A
part of all dead organic matter as well as the exudation of
zooplankton is degraded by bacteria, though the dynamics of the
bacteria themselves is not included in the model. The
phytoplankton grows upon the uptake of nutrients, but its
concentration is diminished by zooplankton (grazing term) and due
to natural mortality. Grazing, modelled by a Holling type III
function, enters also as a growth term for the zooplankton
dynamics multiplied by a factor $\gamma$ taking into account that
only a part of the food is converted into biomass of the
zooplankton, while the other part $(1-\gamma)$ goes to recycling.
The natural mortality of zooplankton is assumed to be quadratic
because this term does not only model natural mortality but also
the existence of higher predators which are not explicitly
considered  \citep{Edwards2001}. The parameters used are taken
from \citep{Pasquero2004} as presented in Table
\ref{bioparameters}. Although appropriate for open ocean, they
would provide estimates for biological properties in the Atlantic
not too close to the coast. To obtain dimensionless quantities
convenient for the numerics, space is measured in units of $r$,
time in units of $T_c$ and mass in units of $10^{12}$ mmol N.

\begin{table}[ht] \label{bioparameters}
\caption{List of parameters used in the biological model}
\begin{tabular}[t]{l l l}
 \hline
parameter & value & dimensionless value\\
 \hline
$\beta$ &  0.66 day$^{-1}$ & 19.8\\
$\eta$ & 1.0 (mmol~N~m$^{-3})^{-2}$~day$^{-1}$ & 0.12288\\
$\gamma$ & 0.75 & 0.75\\
$a$ & 2.0~day$^{-1}$ & 60\\
$S_l$ & 0.00648 day$^{-1}$ (nutrient poor) & 0.1944\\
$S_h$ & 0.648 day$^{-1}$ (nutrient rich)  & 19.44\\
$k_N$ & 0.5 mmol N m$^{-3}$& 7.8125\\
$\mu_N$ & 0.2 & 0.2\\
$\mu_P$ & 0.03 day $^{-1}$ & 0.9\\
$\mu_Z$ & 0.2 (mmol~N~m$^{-3})^{-1}$~day$^{-1}$ & 0.384\\
$N_0$ & 8.0 mmol~N~m$^{-3}$ & 125\\
 \hline
\end{tabular}
\end{table}

The primary production is defined as the growth term in the
phytoplankton dynamics:

\begin{equation}\label{PP}
PP=\beta \frac{N}{k_{N}+N}P
\end{equation}

The function $S$, measuring the strength of vertical mixing in
this model is a crucial quantity for the coupling between the
hydrodynamical and the biological model, because it quantifies the
{\it local} nutrient supply. As shown in Fig. \ref{fig:area} we
assume that there exists an upwelling zone which is located in a
small rectangular region on one side of the island. According to
this assumption, we assign two different values to the parameter
$S(x,y)$. In the upwelling zone there is a strong vertical mixing
leading to nutrient rich waters in the mixed layer. There we
assume $S(x,y)=S_h=0.648$ day$^{-1}$, while in all the surrounding
waters upwelling is much lower so that we assign
$S(x,y)=S_l=0.00648$ day$^{-1}$
 which is a hundred times lower.

The dynamics of this model is different depending on the choice of
parameters. The long-term behavior can be either stationary with
constant concentrations of $N$, $P$ and $Z$ or oscillatory. We
refer for more details to \citet{Edwards1996} and
\citet{Pasquero2004}. We use  a parameter set where the system
possesses a stable steady-state. Using the parameter values from
Table \ref{bioparameters} and setting the vertical mixing
$S=S_l=0.00648$ day$^{-1}$ we obtain as a steady state
$N_{amb}=0.185$, $P_{amb}=0.355$ and $Z=Z_{amb}=0.444$
mmol~N~m$^{-3}$. In this nutrient poor region the ambient primary
production is $PP_{amb}=0.0633$ mmol~N~m$^{-3}$ day$^{-1}$.


\subsection{The coupled model}
\label{Complete model and input conditions}

The coupling of the biology and the hydrodynamics  yields a system
of advection-reaction-diffusion equations. Thus the complete model
is given by the following system of partial differential
equations:

\begin{eqnarray}
\frac{\partial N}{\partial t} + {\vec v}\cdot \nabla N &=& F_N + D
\nabla^2 N,
\nonumber \\
\frac{\partial P}{\partial t} + {\vec v}\cdot \nabla P &=& F_P + D
\nabla^2 P ,
\nonumber \\
\frac{\partial Z}{\partial t} + {\vec v}\cdot \nabla Z &=& F_Z + D
\nabla^2 Z ,
\label{PDE}
\end{eqnarray}

with the biological interactions $F_N$, $F_P$, and $F_Z$ from Eq.
(\ref{BioDLG}), and the velocity field ${\vec v} (x,y,t)$ from
Eqs. (\ref{Strfcomp}) and (\ref{StrfFG}). The diffusion terms take
into account the small scale turbulence with eddy diffusivity $D$.
We take $D\approx 10m^2/s$, as corresponding to the Okubo
estimation of eddy diffusivity at scales of about 10 km
\citep{Okubo1971}, the scales which begin to be missed in our
large scale streamfunction. This advection-reaction-diffusion
system is solved numerically by means of the method explained in
Appendix A. As we are studying an open flow, the inflow conditions
into the left part of the computational domain have to be
specified to complete the model definition. Depending on the
choice of the inflow concentrations we observe different behavior.
A detailed analysis can be found in \citep{Sandulescu2007}, here
we only recall the main results which are the basis of the
analysis we present here. We have studied two essentially
different inflow conditions:

\begin{enumerate}
\item In the first one fluid parcels enter
the computational domain with the ambient concentrations
$N_{amb}$, $P_{amb}$, and $Z_{amb}$ corresponding to the
steady-state for $S=S_l$. In this case the exterior of the
computational domain has the same properties as the part of the
domain without upwelling.

\item In the second one fluid parcels transported by the main flow
enter the domain from the left with a very small content of
nutrients and plankton, corresponding to a biologically very poor
open ocean outside the considered domain. In particular we take
$N_L=0.01\times N_{amb}$, $P_L=0.01\times P_{amb}$ and
$Z_L=0.01\times Z_{amb}$, leading to very low primary production
in the inflow water $PP_L=8.6\times 10^{-6}$ mmol~N~m$^{-3}$
day$^{-1}$. Since those concentrations are very low, we take into
account that fluctuations may be important by adding to each of
the concentrations $(N_L,P_L,Z_L)$ some Gaussian noise term (cf.
\citet{Sandulescu2007} for details).
\end{enumerate}

\begin{figure*} [t]
\begin{center}{
\mbox{\includegraphics[width=\textwidth]{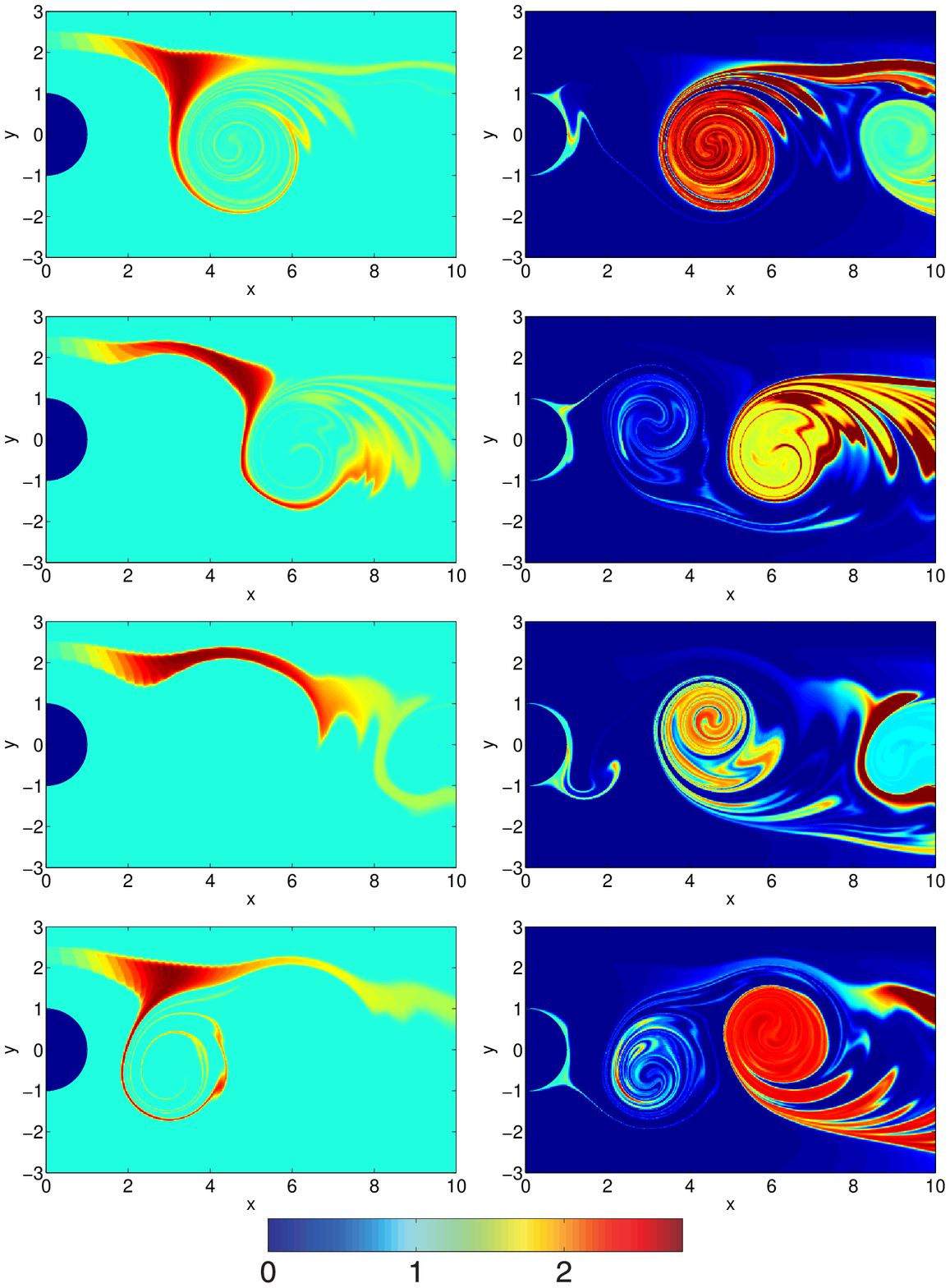}}
}
\end{center}
\caption{
The concentration of phytoplankton (normalized with  the steady
state concentration, $P_{amb}$) for inflow at ambient
concentrations (left) and  for low inflow concentrations (right).
Snapshots taken at $t/T_c=4.1, 4.35, 4.6, 4.85$.
\label{fig:patterns}}
\end{figure*}

The two inflow concentrations yield different behavior as shown in
Fig. \ref{fig:patterns}. In the first case we observe high
biological activity connected with a high primary production in
the area outside vortices. Namely, this is the area of the
nutrient plumes advected from the upwelling region (Fig.
\ref{fig:patterns} left column). By contrast, in the second inflow
case (Fig. \ref{fig:patterns} right column) we obtain a high
phytoplankton concentration within the vortices. It turns out that
here the vortices act as incubators for primary production leading
to localized plankton blooms. In the first inflow case the
behavior is easy to understand since due to higher nutrient
concentrations in the upwelling region and its neighborhood a high
growth of phytoplankton is expected. The response on the upwelling
of nutrients in the second case is less obvious. Therefore the
main objective of this paper is to find out the mechanism of the
localized plankton blooms. In the rest of this work we consider
only the situation that the concentrations at inflow are at their
low values $N_L, P_L, Z_L$.

\section{The mechanism of emergence of localized plankton blooms}
\label{results}

After specifying the complete model and its dynamics we now
investigate the behavior of the coupled biological and
hydrodynamical system from different perspectives to clarify the
mechanism of localized enhancement of phytoplankton and primary
production connected to vortices. Firstly we study the biological
model alone to understand the interplay between the three
biological components $N, P$ and $Z$ leading to a sharp increase
of phytonplankton for some time interval. This study yields a
certain biological time scale for the growth of plankton which we
compare in a second step to the hydrodyamical time scale obtained
from the investigations of residence times in vortices. Thirdly we
discuss the role of the chaotic saddle embedded in the flow for
the emergence of localized enhanced plankton growth.

\subsection{Plankton growth}
\label{Plankton growth}
To study the enhancement of primary production and the emergence
of localized algal blooms we have first to analyze the dynamics of
the biological model. There is no commonly accepted definition of
an algal bloom. Usually a large increase in the phytoplankton
concentrations is considered as a bloom. In most cases such blooms
are observed once or twice a year due to seasonal forcing. In our
case the phytoplankton bloom is not related to an external forcing
and appears only on a rather short time scale. We consider the
case where there appears a sharp increase in phytoplankton as a
result of an enrichment with nutrients \citep{ Edwards1996,
Huppert2002}.

\begin{figure} [t]
\begin{center}{
\includegraphics[width=\columnwidth]{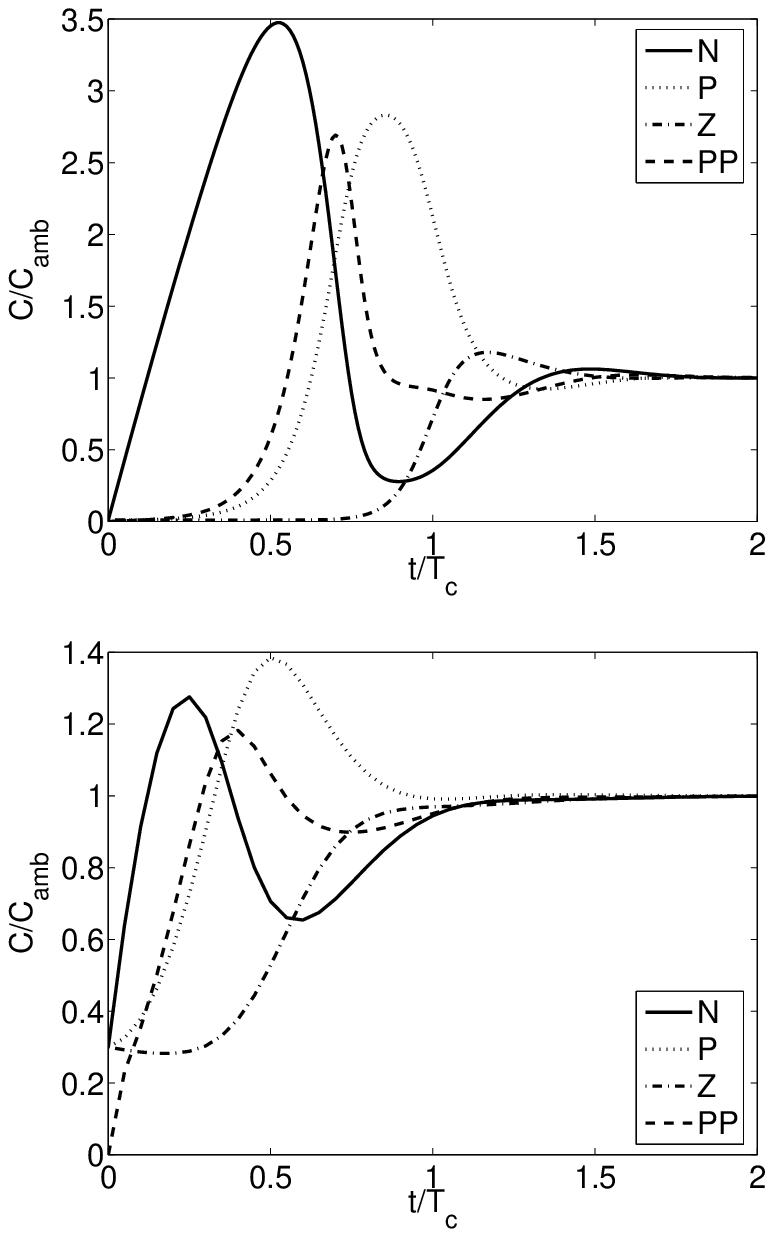}
}
\end{center}
\caption{The time evolution of the concentrations ($C$) of the species and
the primary productivity. Starting concentrations are $0.01\times$
steady state concentrations of $N$, $P$ and $Z$ (top) and
$0.3\times$ steady state concentrations of $N$, $P$ and $Z$
(bottom).
\label{fig:bioalone-over-t}}
\end{figure}

Since the long-term behavior of our model is stationary for the
parametrization used, the emergence of a sharp increase in
phytoplankton is a transient phenomenon and its time scale is
important for the mechanism of localized enhancement of the
primary production. The time evolution of the three components and
the primary production $PP$ of the model system towards the steady
state concentrations $C_{amb}$ ($C=N,P,Z,$ or $PP$) is shown in
the upper part of Fig. \ref{fig:bioalone-over-t}. With starting
concentrations $0.01\times$ steady-state concentrations of
$N_{amb}$, $P_{amb}$ and $Z_{amb}$, first the nutrient
concentration increases and, after a time lag, primary production
and phytoplankton concentration follow with a large increase. This
growth is approximately exponential when the nutrients reach their
maximum. Finally, with a larger time lag the concentration of
predators (zooplankton) increases as well and the bloom ends due
to two factors: nutrient depletion and increased grazing by
zooplankton. For comparison in the lower panel of Fig.
\ref{fig:bioalone-over-t} the time evolution of the system with
starting concentrations $0.3\times$ steady-state concentrations of
$N_{amb}$, $P_{amb}$ and $Z_{amb}$ is plotted. With higher
starting concentrations the overshooting in nutrient and
phytoplankton concentrations at the beginning of the time
evolution is less pronounced (because there are more predators
already present) and the concentrations converge faster towards
the steady-state.

From these simulations we can estimate the time scale for the
biological growth: To reach the maximum of the bloom, only 15 to
25 days are necessary depending on the initial condition. The time
scale for the whole relaxation process is about $2T_c$, i.e. about
60 days. To understand the interplay between the biological growth
and the hydrodynamic mesoscale structures we have to compare this
biological time scale with the hydrodynamic one.

\subsection{The residence time of fluid parcels in the wake}
\label{subs:The residenmce time of parcels in the wake}

As pointed out in \citet{Sandulescu2007} the hydrodynamic
mesoscale structures are important for the enhancement of primary
production in the wake of the island. To gain more insight into
the interplay of hydrodynamics and plankton growth we now quantify
the time scales for the relevant hydrodynamic processes. To this
end we study the various structures in the hydrodynamic flow which
have a significant influence on the residence times of nutrients
and plankton in the wake of the island. Firstly, far away from the
island (top and bottom of Fig. \ref{fig:area}) the flow is strain
dominated and particles like nutrients and plankton are advected
with the background flow of speed $u_0$. Thus the  residence time
of particles released away from the island (with  $y>2$ and
$y<-2$, $x=0$) is about 16 days.

\begin{figure} [t]
\begin{center}{
\includegraphics [angle=0, width=\columnwidth]{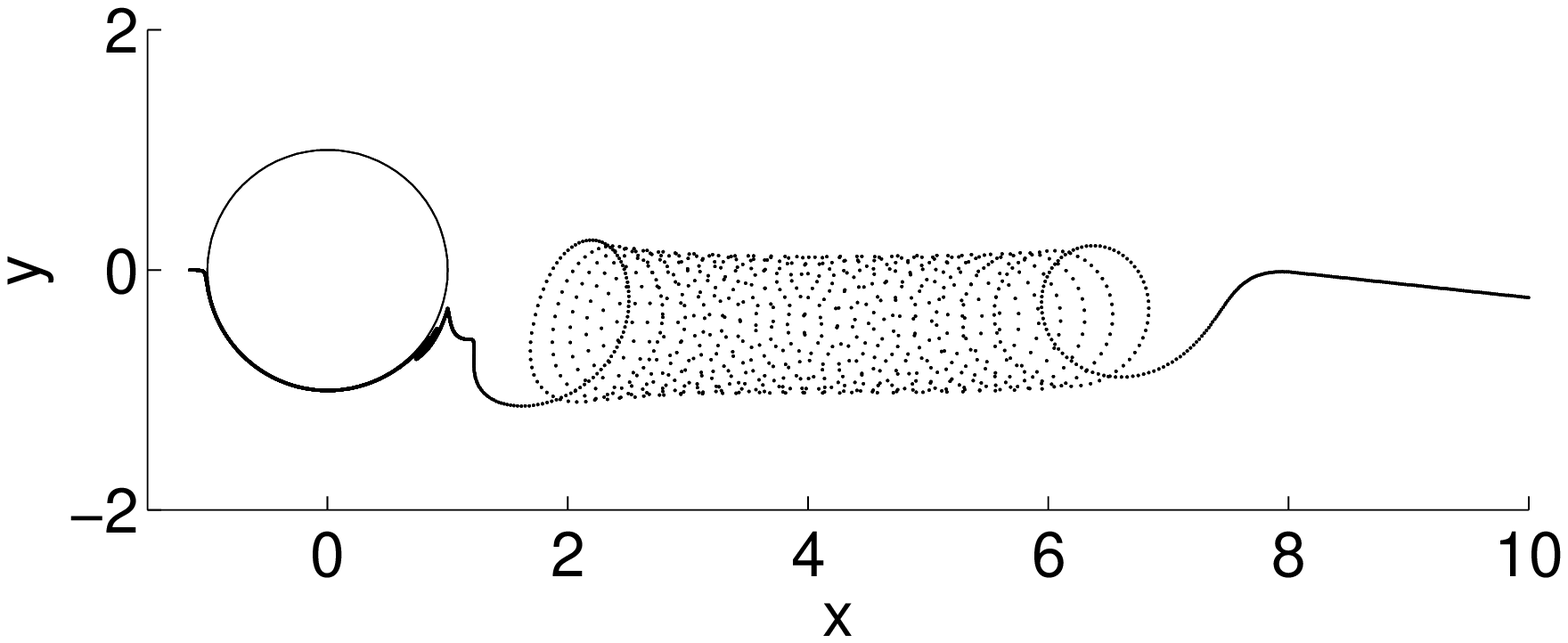}
}
\end{center}
\caption{The trajectory of a fluid parcel released in the flow at the
coordinates (-1.15, 0) at time $t=0$. Its subsequent positions are
plotted with a dimensionless time step $\Delta t=0.001$.
\label{fig:trajectory}} \end{figure}

Secondly we note the existence of the eddies. They are
characterized by a dominance of vorticity compared to strain. Thus
particles are trapped in the vortex once entrained to it. The
particles will rotate in the vortex for some time, but since this
confinement is not perfect and vortices exist only for some time
they leave the vortex and move away with the background flow out
of the computational area (cf. Fig. \ref{fig:trajectory}).

\begin{figure} [t]
\begin{center}{
\includegraphics[width=0.45\textwidth]{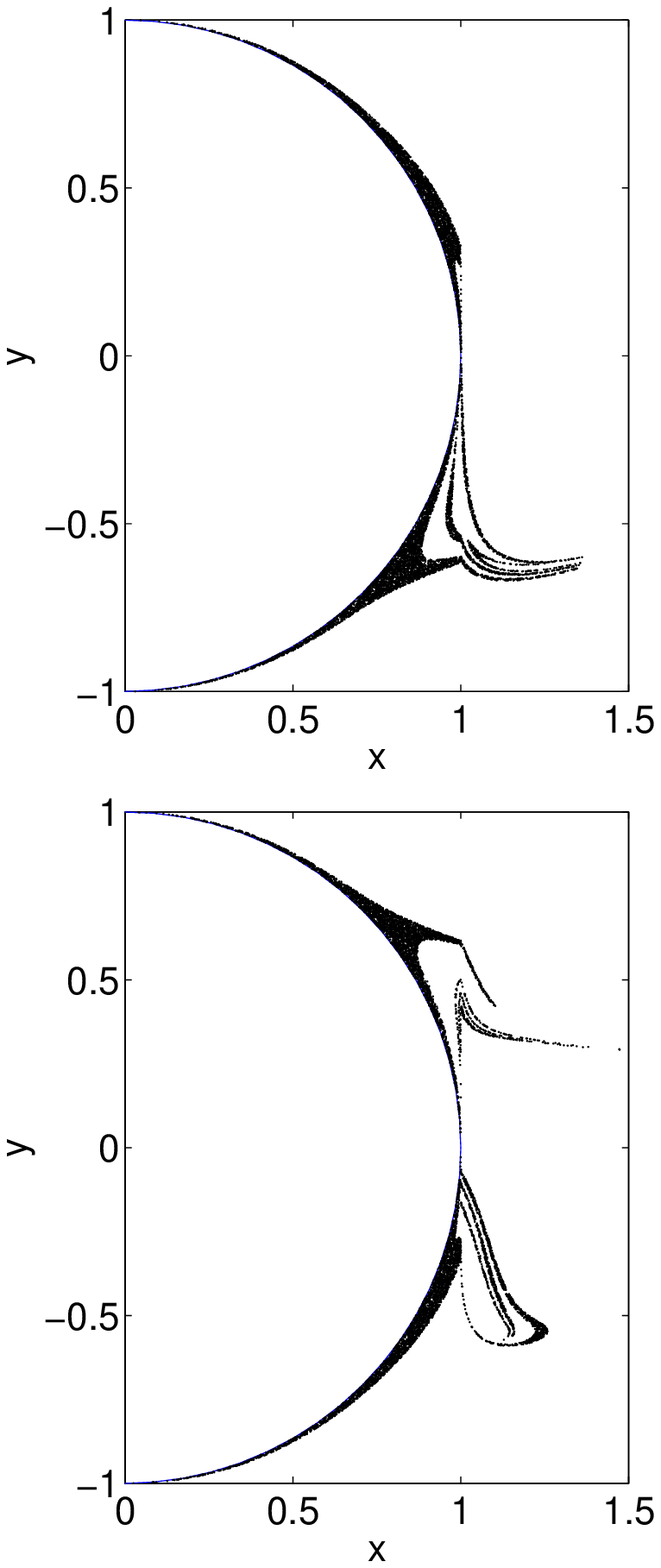}
}
\end{center}
\caption{
The unstable manifold of the chaotic saddle. Snapshots taken at
$t=1.5 \ T_c$ (lower plot), and $ 2 \ T_c$ (upper plot).
\label{fig:unstable-mani}}
\end{figure}

\begin{figure} [t]
\includegraphics [width=\columnwidth]{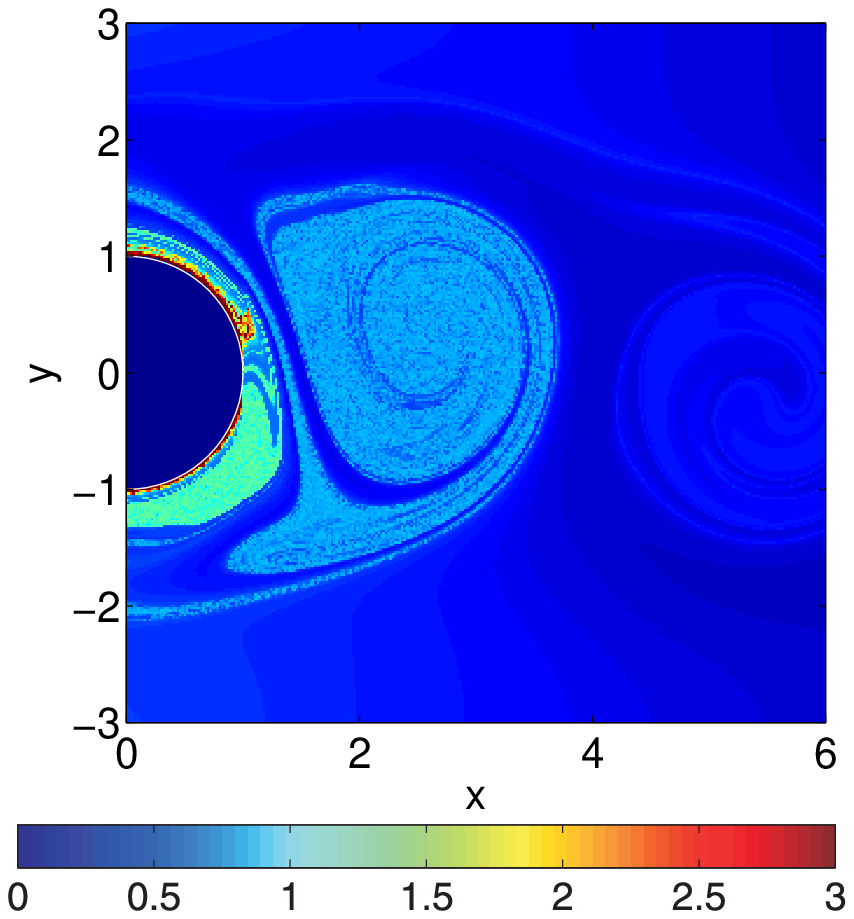}
\caption{Residence time (coded in color in
units of $T_c$) of fluid elements inside the area as a function of
its initial positions at time $t/T_{c}=0.25$.
\label{fig:res-time}}
\end{figure}

\begin{figure*} [t]
\begin{center}{
\includegraphics[width=\textwidth]{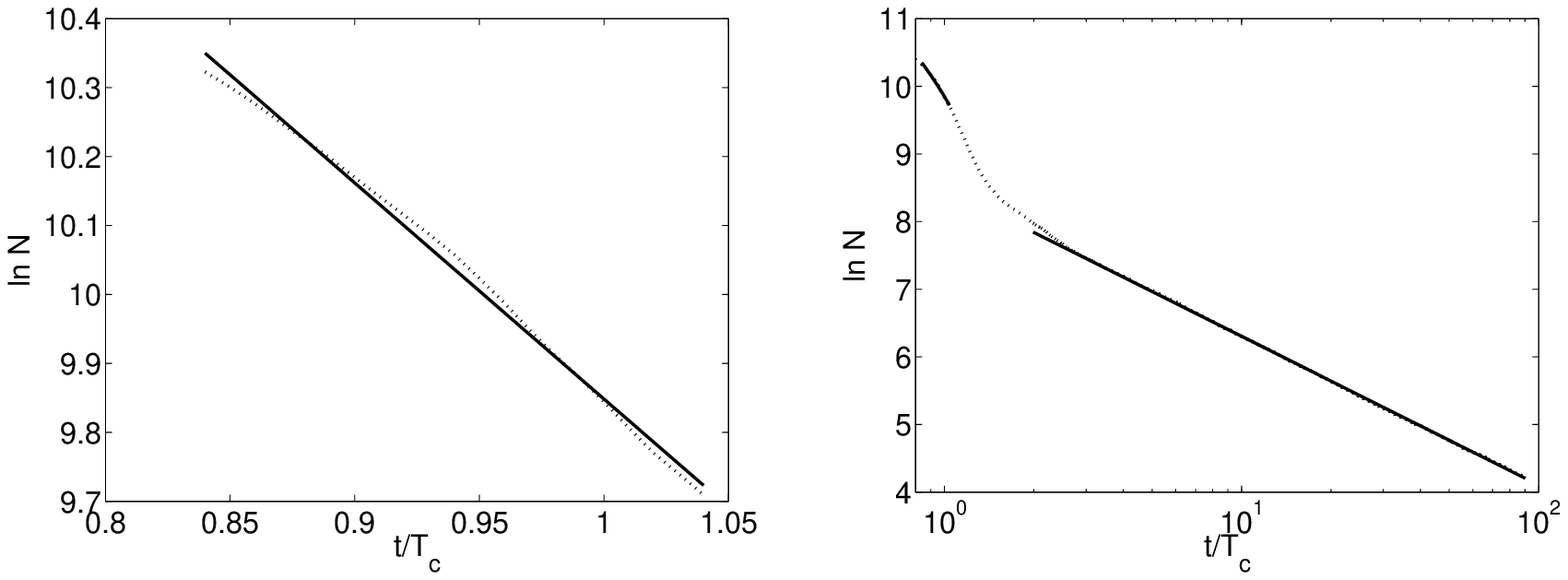}
}
\end{center}
\caption{The number $N$ of tracers inside the area $[0,2]\times[-1,1]$ as a function
of time in units of the period of the flow $T_c$. Vortex strength
$w=200$. Left panel shows the very early decay, with an
exponential fit. Right panel plots the overall decay, with a power
law fit at large times.
\label{fig:res-time-fit}} \end{figure*}

Thirdly we consider two other geometrical objects which are also
relevant for the residence time of particles in the vortex street:
the chaotic saddle and the cylinder boundary. As shown in
\citep{Jung1993,Duan1997} there exists a chaotic saddle which is
embedded in the flow beyond the island. At least for short time
scales, this invariant set determines the residence times of
particles. Particles released in the neighborhood of the chaotic
saddle will approach it along its stable manifold and will leave
it along its unstable manifold.
 The unstable manifolds
at two different times (the manifolds and the saddle are
time-dependent) are presented in Fig. \ref{fig:unstable-mani}. As
compared with the configuration in \citet{Jung1993}, the manifolds
are tightly packed close to the cylinder, because of the
parameters used here. To obtain an estimate for the residence time
on the chaotic saddle we use a method suggested by
\citet{Jung1993}. We sprinkle a large number $N$ of tracer
particles ($N=62500$) in the area $[0,2]\times[-1,1]$ and
integrate their trajectories forward in time. If the dynamics in
the region is mainly hyperbolic, the number of particles remaining
in the area of the chaotic saddle decreases as
$N~\sim~\exp(-\kappa~t)$ with $\kappa$ the escape rate or
$1/\kappa$ the mean residence time on the saddle. Figure
\ref{fig:res-time} shows the residence times obtained with this
method, and Fig. \ref{fig:res-time-fit} shows the decay of the
number of particles in the region as a function time. We note that
the expected exponential decay occurs only for very short time
scales. By fitting this initial time decay, the corresponding
escape rate is $\kappa=3.1/T_c$, and therefore the residence time
of tracers in the hyperbolic part of the saddle is $\tau~\sim~10$
days. For larger times the particle number in the region decays as
a power law. The reason for this power-law behavior is the
non-hyperbolic dynamics near the boundary of the cylinder. As
already shown by \citep{Jung1993} particles stay for a long time
in the vicinity of the cylinder giving rise to another long-term
statistics of the residence times of the tracers. Thus the number
of particles decays as $N \sim t^{-\gamma} $ with $\gamma=0.96$.
The residence times in the vicinity of the island can be estimated
as $\tau~\sim~85$ days, measured from the decay to a fraction $e$
of the initial number: $N(\tau)=N(0)/e$.

Overall we obtain a residence time statistics which reflects a
combination of the three components in the flow: the cylinder, the
chaotic saddle and the vortices. This leads to residence times of
tracers in the wake as long as $90$ days (cf.
Fig.~\ref{fig:res-time}). Note that in Fig. \ref{fig:res-time} one
can see that tracers having the longer residence times are either
located close to the cylinder or on the chaotic saddle. The
residence times in the vortices are determined by their travel
time which is about 50 days.

\subsection{The interplay of biological and hydrodynamical residence
times}
\label{The interplay of biological and hydrodynamical residence
times}

To understand the emergence of localized enhancement of primary
production we have to analyze the interplay of the different time
scales relevant for coupled biological and hydrodynamical
processes. Biological evolution needs about 30-60 days to reach
the steady-state when entering the computational domain with very
low concentrations of nutrients and plankton. Due to the
exponential growth in the beginning of the growth phase, we obtain
a plankton bloom after about 25 days. Outside the vortex street
the travel time of tracers through the computational domain is
only about 16 days due to the background flow of $u_0=0.18$ m/s.
Therefore we cannot expect a considerable growth of plankton
outside the vortex street, since the residence time of plankton
and nutrients is too short.

Let us now analyze the situation within the mesoscale structures
of the flow in the wake of the island. As the residence time close
to the island is about 85 days the concentrations of nutrients and
plankton have already reached the steady-state which is also
indicated by the green color in the right column of Fig.
\ref{fig:patterns}. Some of the particles in the vicinity of the
island come close to the stable manifold of the chaotic saddle
visible as the filaments which detach from the cylinder. These
filaments are stretched and folded along the unstable manifold of
the chaotic saddle, being diffusively diluted during the process
by mixing with the poor surrounding waters. Thus, very thin
filaments of low plankton and nutrient concentration are produced
which are first rolled around the vortices and then entrained by
them. Inside the vortex the concentrations become homogeneised to
a low value.
 These very low concentrations of plankton experience
the bloom cycle described in Section \ref{Plankton growth} during
the time they are trapped and confined by the vortex. Since the
travel time for the vortices is about 50 days, plankton in them
has time to grow. Therefore we observe a localized plankton bloom
when the vortex has travelled a distance of $\sim 100$ km which
corresponds to a residence of the plankton in the vortices of
$\sim 15-20$ days. After 40-50 days and $\sim 200$ km we obtain
steady-state concentrations and the former filamental structure
within the vortex is smeared out by our diffusion term which
mimics small scale turbulence.

\subsection{The emergence of filamental structures due to strong mixing}
\label{The emergence of filamental structures due to strong mixing}

In the previous subsection we have stated that the transport of
nutrients and plankton from the vicinity into the interior of a
vortex happens by filaments which are entrained by the vortex. To
explain this stretching mechanism we now study the mixing process
around the vortices in more detail using a method to visualize
exponential divergence of the trajectories of initially nearby
particles.

The usual tool to analyse exponential divergence in dynamical
systems theory is the computation of Lyapunov exponents. In order
to adjust this concept to local processes, we compute finite size
Lyapunov exponents (FSLE) which are based on the idea that one
measures the time necessary to obtain a final prescribed distance
$\delta_f$ starting from an initial distance $\delta_0$
\citep{Artale1997,DOvidio2004}. For a two-dimensional flow we
obtain two Lyapunov exponents $\lambda_+$ and $\lambda_-$ (see
appendix B).

\begin{figure} [t]
\begin{center}{
\includegraphics[width=\columnwidth]{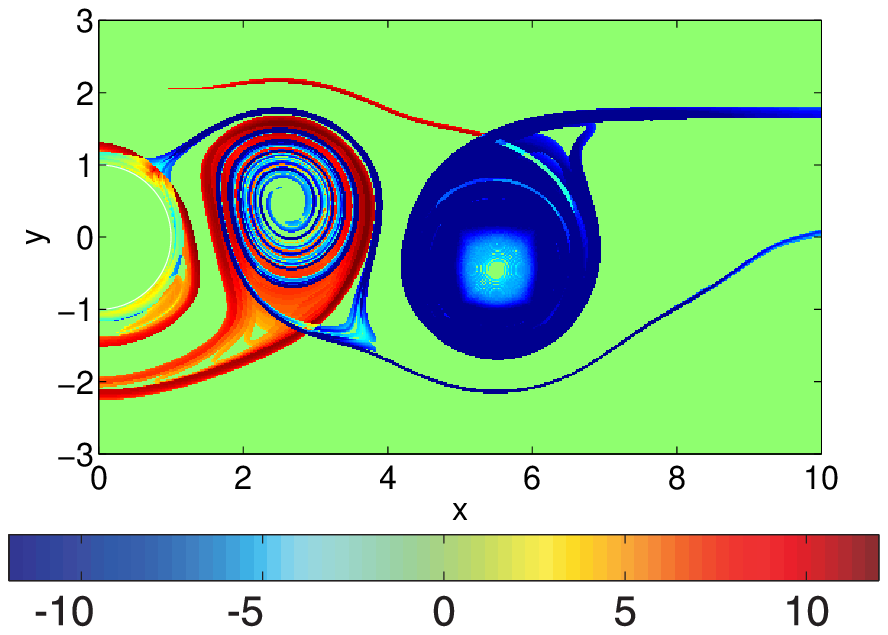}
}
\end{center}
\caption{Finite Size Lyapunov Exponent distributions. We plot the field $\lambda_{+}
-|\lambda_{-}|$ at time $t/T_{c}=0.25$. Stable and unstable
manifolds in the flow are approximated by the most positive and
most negative filaments in the distributions.
\label{fig:fsle-v-r}}
\end{figure}

Maxima in the spatial distribution of $\lambda_+$, the positive or
expanding FSLE, approximate the underlying stable manifold of the
chaotic flow \citep{Joseph2002,DOvidio2004}, the direction along
which parcels approach the saddle. The contracting FSLE,
$\lambda_-$, detects the underlying unstable manifold of the
chaotic flow, the direction along which parcels are stretched out
of the saddle. For details on how to calculate these scalar fields
see Appendix B.

The FSLEs were calculated choosing the initial separation
$\delta_0$ equal to the gridsize and the final separation
$\delta_f$ equal to the radius of the island and the vortices,
since this is the scale of the motion in the wake. As both stable
and unstable manifolds cannot be crossed by fluid parcels they are
barriers \citep{Artale1997,DOvidio2004}. The scalar field
$\lambda_+ - |\lambda_-|$ is plotted in Fig. \ref{fig:fsle-v-r}.
FSLEs are Lagrangian measures, which are computed from
trajectories that remain in the flow for a long time, in our case
for up to $3T_c$. Therefore even though they are plotted as a
snapshot, the visualized structures reflect the stretching and
folding of the fluid parcels during this long time.
The stable and unstable manifolds are intertwined around the
vortex cores and at the island. Stable and unstable manifolds are
crossing the wake allowing for transport across the vortex street.
They intersect each other in hyperbolic points, regions of strong
mixing. This stretching-compressing mechanism leads to low
nutrient and plankton concentrations transported into the interior
of the vortex, and thus becoming the starting concentrations for
the localized plankton bloom.

\subsection{On the role of the upwelling region of nutrients}
\label{On the role of the upwelling region for nutrients}

\begin{figure} [t]
\begin{center}{
\includegraphics[width=.7\textwidth]{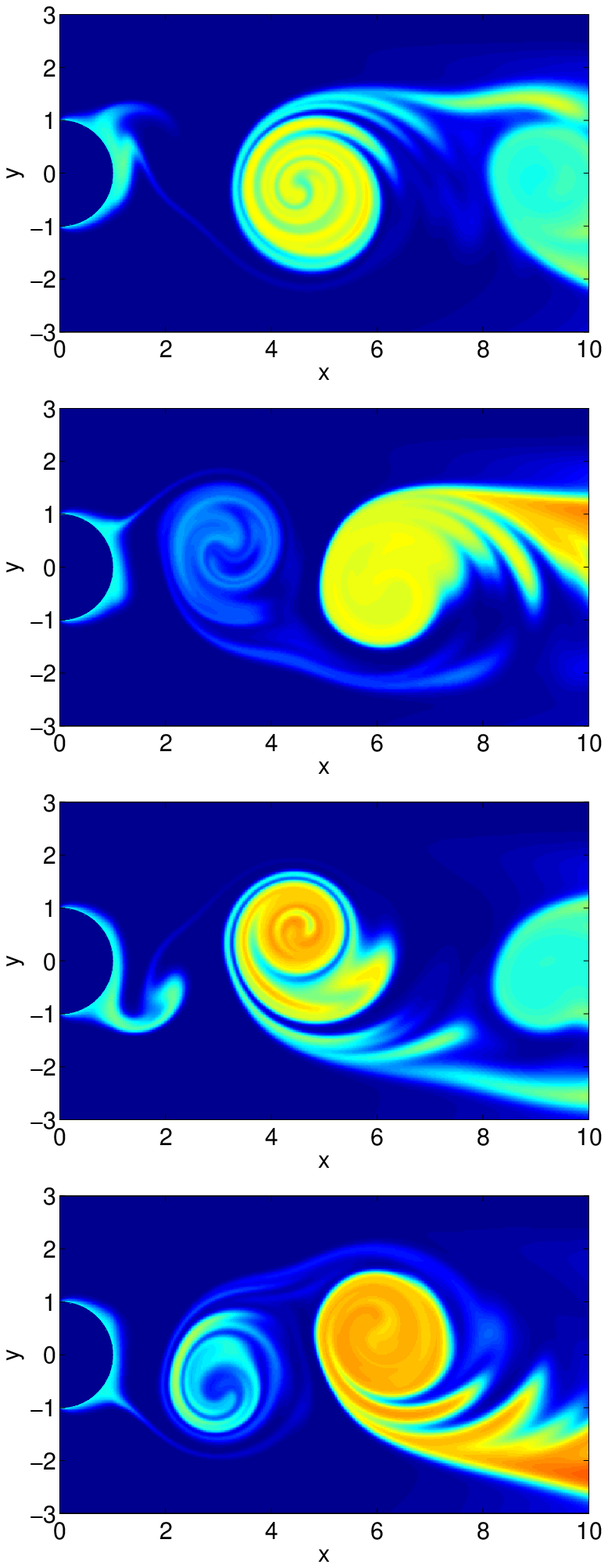}
}
\end{center}
\caption{
The concentration of phytoplankton in the absence of upwelling.
Snapshots taken at $t/T_c=4.1, 4.35, 4.6, 4.85$. The color coding
is in the same range as in Fig. \ref{fig:patterns}.
\label{fig:patterns-noupwelling}}
\end{figure}

Finally we discuss the importance of the vertical mixing of
nutrients in the upwelling zone for the emergence of a plankton
bloom inside vortices. Comparing Fig. \ref{fig:patterns} left and
right column it is obvious that in the case of an inflow with
steady-state conditions (left column), the nutrient plume which
appears in the neighborhood of the upwelling zone gives rise to a
phytoplankton bloom (red filamental plume). Such a plume is almost
absent under low inflow conditions (right column).
 Though the nutrient supply due to vertical mixing is identical for
both inflow conditions, it seems to have a limited effect in the
low inflow case. One argument has been already discussed above:
The background flow transports the nutrients too fast so that the
very small plankton concentrations can not grow to reach high
values during the travel time through the computational area. The
growth of phytoplankton is visible only further downstream. This
leads to the conclusion that the plankton bloom inside the vortex
is only slightly influenced by the extra nutrients entrained from
the upwelling zone in the low inflow situation.
 To strengthen this statement we present in Fig.
\ref{fig:patterns-noupwelling} the plankton dynamics when the
upwelling is removed. We note that the concentration values for
phytoplankton and zooplankton are slightly lower compared to the
upwelling regime, but qualitatively there is no change observable.
Thus localized phytoplankton blooms in vortices are possible in
the wake of an island just due to the mechanism discussed in
Subsec. \ref{The emergence of filamental structures due to strong
mixing} without any extra nutrient supply due to upwelling.

\section{Conclusions}
\label{conclusions}

We have analyzed the interplay between hydrodynamic mesoscale
structures and biological growth in the wake of an island.
Parameter values for the kinematic hydrodynamic flow were chosen
to match the observations for the Canary island region, but since
the basic hydrodynamic features studied here are commonly observed
in other areas too, we expect our results to be of general
validity. Our study is focused on the emergence of a plankton
bloom localized in a vortex in the wake of an island. In a
previous paper \citep{Sandulescu2007} it has been pointed out that
under certain conditions a vortex may act as an incubator for
plankton growth and primary production. Here we have revealed the
mechanism of such a plankton bloom. If the hydrodynamic flow far
away from the island is dominated by a jet, then the hydrodynamic
time scale is much faster than the biological one, so that
considerable growth of plankton cannot be observed. By contrast,
in the wake of an island we obtain a much slower time scale which
becomes comparable to the biological one giving rise to an
exponential growth of phytoplankton and thus to the emergence of a
plankton bloom within a vortex. The essential factors for this
phenomenon to happen are (i) the long residence times in the
vicinity of the island leading to an enrichment of nutrients and
plankton in the neighborhood of the island; (ii) the transport and
subsequent entrainment of nutrients and plankton to the interior
of the vortex due to filamental structures emerging with the
chaotic saddle beyond the island, and (iii) the confinement of
plankton in the vortex. Though the upwelling of nutrients in an
upwelling zone enhances the emergence of localized plankton
blooms, it is not a precondition for this phenomenon to occur. The
extra nutrients supplied by vertical mixing in areas away from the
vortex street are not a part of the mechanism explained here.
Upwelling could be more effective if the vortices directly cross
upwelling zones when travelling through the ocean. Similar
situations have been considered in \citep{Martin2002,
Pasquero2005}. There it has been shown that under conditions where
upwelling occurs only directly within vortices, a plankton bloom
within a vortex can be initiated. Therefore this mechanism, which
relies mostly on upwelling, is different from the one reported
here. The variety of real observations \citep{Aristegui1997} in
the Canary wake may benefit from the identification of the
different possible mechanisms.


\section*{Appendix A: The numerical algorithm}
\label{algorithm}

The investigation of the interplay of biological and physical
processes is based on advection-reaction-diffusion systems
(Eqs.~\ref{PDE}). This system of partial differential equations is
solved numerically by means of a semi-Lagrangian algorithm. The
concentration fields of nutrients $N$, phytoplankton $P$ and
zooplankton $Z$ are represented on a grid of [500 x 300] points.
The integration scheme splits the computation into three steps
corresponding to advection, reaction and diffusion which are
performed sequentially in the following way:
\begin{enumerate}

\item Advection: Each point of the grid is integrated for a time
step $dt$ backwards in time along the trajectory of a fluid parcel
in the velocity field. This procedure yields the position from
which a fluid parcel would have reached the chosen grid point.
Typically this position is not located on a grid point but
somewhere in between.

\item Reaction: Once the position of the fluid parcel in the past
is found, we compute the values of the concentration fields of
$N$, $P$ and $Z$ at this point and take them as initial values for
the reaction term which is integrated forward in time for a time
step $dt$. Since the position of the fluid parcel is not on a grid
point the concentration fields have to be evaluated by means of a
bilinear interpolation using the nearest neighbor grid points.

\item Diffusion: Finally we perform a
diffusion step based on an Eulerian scheme. Note, that the
reaction step induces already a numerical diffusion of the order
$D_n\propto dx^2/dt$ due to the interpolation. Therefore one has
to make  sure that the real diffusion according to the Okubo
estimate (~$D=10m^2/s$, dimensionless value $D=0.041472$)
\citep{Okubo1971} is larger than this numerical diffusion.
Additionally the stability condition of the Eulerian diffusion
step ($D dt_d /dx^2 < 1$ with $dt_d$ the diffusion time step) has
to be fulfilled. Both conditions together require that the
diffusion time step $dt_d$ is much smaller than $dt$. We have
chosen $dt_d=dt/10$ in our algorithm, i.e. after each step of
advection and reaction we perform 10 steps of diffusion. The
parameters used in the computation are $dx=0.02$, $dt=0.01$ and
$dt_d=0.001$ expressed in units of $T_c=30~days$ for time and
$r=25~km$ for space.

\end{enumerate}

\section*{Appendix B: Finite Size Lyapunov-Exponents}
\label{Appendix: FSLE}

Stretching by advection in fluid flows is often described by means
of Lyapunov exponents. They are defined as the average of the
exponential rate of separation of initially infinitesimally
separated parcels. For application with data sets from tracer
experiments the infinite time limit poses a problem. To study
non-asymptotic dispersion processes, Finite Size Lyapunov
Exponents (FSLE) have been introduced
\citep{Artale1997,DOvidio2004}. The FSLE technique allows us to
characterize dispersion processes and to detect and visualize
Lagrangian structures, such as barriers and vortices. The FSLE are
computed by starting two fluid elements at time $t$ close to the
point ${\bf x}$ but at a small distance $\delta_0$, and let them
to evolve until their separation exceeds $\delta_f$. From the
elapsed time, $\tau_+$, the FSLE is calculated as
\begin{equation}
\label{FSLE+def}
\lambda_{+}({\bf x}, t, \delta_0, \delta_f)= \frac{1}{\tau_+} \log
\frac{\delta_f}{\delta_0}
\end{equation}
The positive subindexes indicate that the tracers are advected
forward in time. $\lambda_{+}$ is a scalar measure giving the
stretching rate in the flow as it is the inverse of the separation
time $\tau$.

The same definition can be applied to tracers integrated in the
negative direction in time. $\lambda_{-}$ gives the contraction
rate in the flow at the specified position:
\begin{equation}
\label{FSLE-def}
\lambda_{-}({\bf x}, t, \delta_0, \delta_f)= \frac{1}{\tau_-} \log
\frac{\delta_f}{\delta_0}
\end{equation}

Regions with high values of $\lambda_{+}$ and $\lambda_{-}$ trace
out approximately the stable and unstable manifolds of the chaotic
saddle. These manifolds cannot be crossed by fluid parcel
trajectories and thus greatly influence the transport in the area.

\section*{Acknowledgements} The authors thank T. T{\'e}l for many
inspiring discussions.
 M.S. and U.F. acknowledge financial
support by the DFG grant FE 359/7-1. E.H-G. and C.L. acknowledge
financial support from MEC (Spain) and FEDER through project
CONOCE2 (FIS2004-00953), and PIF project OCEANTECH from Spanish
CSIC. Both groups have benefitted from a MEC-DAAD joint program.


\begin{thebibliography}{32}
\providecommand{\natexlab}[1]{#1} \providecommand{\url}[1]{{\tt
#1}} \providecommand{\urlprefix}{} \expandafter\ifx\csname
urlstyle\endcsname\relax
  \providecommand{\doi}[1]{doi:\discretionary{}{}{}#1}\else
  \providecommand{\doi}{doi:\discretionary{}{}{}\begingroup
  \urlstyle{rm}\Url}\fi

\bibitem[{Abraham(1998)}]{Abraham1998}
Abraham, E.: The generation of plankton patchiness by turbulent
stirring,
  Nature, 391, 577--580, 1998.


\bibitem[{Ar{\'\i}stegui et~al.(1997)Ar{\'\i}stegui, Tett,
  Hern{\'a}ndez-Guerra, Basterretxea, Montero, Wild, Sangr{\'a},
  Hern{\'a}ndez-Leon, Canton, Garc{\'\i}a-Braun, Pacheco, and
  Barton}]{Aristegui1997}
Ar{\'\i}stegui, J., Tett, P., Hern{\'a}ndez-Guerra, A.,
Basterretxea, G.,
  Montero, M.~F., Wild, K., Sangr{\'a}, P., Hern{\'a}ndez-Leon, S., Canton, M.,
  Garc{\'\i}a-Braun, J.~A., Pacheco, M., and Barton, E.: The influence of
  island-generated eddies on chlorophyll distribution: a study of mesoscale
  variation around Gran Canaria, Deep Sea Research I, 44, 71--96, 1997.

\bibitem[{Ar{\'\i}stegui et~al.(2004)Ar{\'\i}stegui, Barton, Tett, Montero,
  Garc{\'\i}a-Mu{\~n}oz, Basterretxea, Cussatlegras, Ojeda, and
  de~Armas}]{Aristegui2004}
Ar{\'\i}stegui, J., Barton, E.~D., Tett, P., Montero, M.~F.,
  Garc{\'\i}a-Mu{\~n}oz, M., Basterretxea, G., Cussatlegras, A.-S., Ojeda, A.,
  and de~Armas, D.: Variability in plankton community structure, metabolism,
  and vertical carbon fluxes along an upwelling filament (Cape Juby, NW
  Africa), Progress in Oceanography, 62, 95--113, 2004.

\bibitem[{Artale et~al.(1997)Artale, Boffetta, Celani, Cencini, and
  Vulpiani}]{Artale1997}
Artale, V., Boffetta, G., Celani, M., Cencini, M., and Vulpiani,
A.: Dispersion
  of passive tracers in closed basins: Beyond the diffusion coefficient, Phsy.
  Fluids, 9, 3162--3171, 1997.

\bibitem[{Baretta et~al.(1997)Baretta, Baretta, and Ebenh\"oh}]{Baretta1997}
Baretta, J., Baretta, J., and Ebenh\"oh, W.: Microbial dynamics in
the marine
  ecosystems model ERSEM II with decoupling carbon assimilation and nutrient
  uptake, JSR, 38, 195 -- 212, 1997.


\bibitem[{Bracco et~al.(2000)Bracco, Provenzale, and Scheuring}]{Bracco2000}
Bracco, A., Provenzale, A., and Scheuring, I.: Mesoscale vortices
and the
  paradox of the plankton, Proc. Roy. Soc. Lond. B, 267, 1795--1800, 2000.

\bibitem[{Denman and Gargett(1995)}]{Denman1995}
Denman, K. and Gargett, A.: Biological-physical interactions in
the upper
  ocean: the role of vertical and small scale transport processes, Annu. Rev.
  Fluid Mech., 27, 225--255, 1995.

\bibitem[{d'Ovidio et~al.(2004)d'Ovidio, Fern{\'a}ndez,
  Hern{\'a}ndez-Garc{\'\i}a, and L{\'o}pez}]{DOvidio2004}
d'Ovidio, F., Fern{\'a}ndez, V., Hern{\'a}ndez-Garc{\'\i}a, E.,
and L{\'o}pez, C.: Mixing structures in the Mediterranean Sea from
finite-size Lyapunov
  exponents, Geophysical Research Letters, 31, L17\,203, 2004.

\bibitem[{Duan and Wiggins(1997)}]{Duan1997}
Duan, J. and Wiggins, S.: Lagrangian transport and chaos in the
near wake of
  the flow around an obstacle: a numerical implementation of lobe dynamics,
  Nonlin. Proc. Geophys., 4, 125--136, 1997.

\bibitem[{Edwards and Bees(2001)}]{Edwards2001}
Edwards, M. and Bees, M.: Generic dynamics of a simple plankton
population
  model with a non-integer exponent of closure, Chaos, Solitons \& Fractals,
  12, 289, 2001.

\bibitem[{Edwards and Brindley(1996)}]{Edwards1996}
Edwards, M. and Brindley, J.: Oscillatory behavior in a
three-component
  plankton population model, Dyn. Stab. Sys., 11, 347--370, 1996.

\bibitem[{Hern\'{a}ndez-Garc\'{\i}a and L\'{o}pez(2004)}]{Hernandez2004}
Hern\'{a}ndez-Garc\'{\i}a, E. and L\'{o}pez, C.: Sustained
plankton blooms
  under open chaotic flows, Ecol. Complex., 1, 253--259, 2004.

\bibitem[{Hern\'{a}ndez-Garc\'{\i}a et~al.(2002)Hern\'{a}ndez-Garc\'{\i}a,
  L\'{o}pez, and Neufeld}]{Hernandez2002}
Hern\'{a}ndez-Garc\'{\i}a, E., L\'{o}pez, C., and Neufeld, Z.:
Small-scale
  structure of nonlinearly interacting species advected by chaotic flows,
  Chaos, 12, 470--480, 2002.

\bibitem[{Hern{\'a}ndez-Garc{\'\i}a et~al.(2003)Hern{\'a}ndez-Garc{\'\i}a,
  L{\'o}pez, and Neufeld}]{Hernandez2003b}
Hern{\'a}ndez-Garc{\'\i}a, E., L{\'o}pez, C., and Neufeld, Z.:
Spatial Patterns
  in Chemically and Biologically Reacting Flows, in: Chaos in Geophysical
  Flows, edited by Bofetta, G., Lacorata, G., Visconti, G., and Vulpiani, A.,
  OTTO Editore, Torino, 2003.

\bibitem[{Huppert et~al.(2002)Huppert, Blasius, and Stone}]{Huppert2002}
Huppert, A., Blasius, B., and Stone, L.: A model of phytoplankton
blooms,
  American Naturalist, 159, 156--171, 2002.

\bibitem[{Joseph and Legras(2002)}]{Joseph2002}
Joseph, B. and Legras, B.: Relation between kinematic boundaries,
stirring, and
  barriers for the Antartic Polar vortex, J. Atm. Sci., 59, 1198--1212, 2002.

\bibitem[{Jung et~al.(1993)Jung, T\'el, and Ziemniak}]{Jung1993}
Jung, C., T\'el, T., and Ziemniak, E.: Application of scattering
chaos to
  particle transport in a hydrodynamical flow, Chaos, 3, 555--568, 1993.

\bibitem[{K\'arolyi et~al.(2000)K\'arolyi, P\'entek, Scheuring, T\'el, and
  Toroczkai}]{Karolyi2000}
K\'arolyi, G., P\'entek, A., Scheuring, I., T\'el, T., and
Toroczkai, Z.:
  Chaotic flow: the physics of species coexistence, Proc. Natl. Acad. Sci. USA,
  97, 13\,661--13\,665, 2000.

\bibitem[{L\'{o}pez et~al.(2001{\natexlab{a}})L\'{o}pez,
  Hern\'{a}ndez-Garc\'{\i}a, Piro, Vulpiani, and Zambianchi}]{Lopez2001b}
L\'{o}pez, C., Hern\'{a}ndez-Garc\'{\i}a, E., Piro, O., Vulpiani,
A., and
  Zambianchi, E.: Population dynamics advected by chaotic flows: A
  discrete-time map approach, Chaos, 11, 397--403, 2001{\natexlab{a}}.

\bibitem[{L\'{o}pez et~al.(2001{\natexlab{b}})L\'{o}pez, Neufeld,
  Hern\'{a}ndez-Garc\'{\i}a, and Haynes}]{Lopez2001}
L\'{o}pez, C., Neufeld, Z., Hern\'{a}ndez-Garc\'{\i}a, E., and
Haynes, P.:
  Chaotic advection of reacting substances: Plankton dynamics on a meandering
  jet, Phys. Chem. Earth, B 26, 313--317, 2001{\natexlab{b}}.

\bibitem[{Mann and Lazier(1991)}]{Mann1991}
Mann, K. and Lazier, J.: Dynamics of marine ecosystems.
Biological-physical
  interactions in the oceans, Blackwell Scientific Publications, Boston, 1991.

\bibitem[{Martin(2003)}]{Martin2003}
Martin, A.: Phytoplankton patchiness: the role of lateral stirring
and mixing,
  Progress in Oceanography, 57, 125--174, 2003.

\bibitem[{Martin et~al.(2002)Martin, Richards, Bracco, and
  Provenzale}]{Martin2002}
Martin, A., Richards, K., Bracco, A., and Provenzale, A.: Patchy
productivity
  in the open ocean, Global Biogeochemical Cycles, 16, 10.1029/2001GB001\,449,
  2002.

\bibitem[{Okubo(1971)}]{Okubo1971}
Okubo, A.: Oceanic diffusion diagrams, Deep-Sea Res., 18, 789,
1971.

\bibitem[{Oschlies and Gar\c{c}on(1999)}]{Oschlies1999}
Oschlies, A. and Gar\c{c}on, V.: An eddy-permitting coupled
physical-biological
  model of the North-Atlantic, sensitivity to advection numerics and mixed
  layer physics, Global Biocheochem. Cycles, 13, 135--160, 1999.

\bibitem[{Pasquero et~al.(2004)Pasquero, Bracco, and Provenzale}]{Pasquero2004}
Pasquero, C., Bracco, A., and Provenzale, A.: Coherent vortices,
Lagrangian
  particles and the marine ecosystem, in: Shallow Flows, edited by Uijttewaal,
  W. and Jirka, G., pp. 399--412, Balkema Publishers, Leiden, 2004.

\bibitem[{Pasquero et~al.(2005)Pasquero, Bracco, and Provenzale}]{Pasquero2005}
Pasquero, C., Bracco, A., and Provenzale, A.: Impact of
spatiotemporal viariability of the nutrient flux on primary
productivity in the ocean, J. Geophys. Res. {\bf 11}, 1--13, 2005.


\bibitem[{Peters and Marras{\'e}(2000)}]{Peters2000}
Peters, F. and Marras{\'e}, C.: Effects of turbulence on plankton:
an overview
  of experimental evidence and some theoretical considerations, Mar. Ecol.
  Prog. Ser., 205, 291--306, 2000.

\bibitem[{Sandulescu et~al.(2006{\natexlab{a}})Sandulescu,
  Hern{\'a}ndez-Garc{\'\i}a, L{\'o}pez, and Feudel}]{Sandulescu2006}
Sandulescu, M., Hern{\'a}ndez-Garc{\'\i}a, E., L{\'o}pez, C., and
Feudel, U.:
  Kinematic studies of transport across an island wake, with application to the
  {C}anary islands, Tellus A, 58, 605--615, 2006{\natexlab{a}}.

\bibitem[{Sandulescu et~al.(2007{\natexlab{b}})Sandulescu,
  Hern{\'a}ndez-Garc{\'\i}a, L{\'o}pez, and Feudel}]{Sandulescu2007}
Sandulescu, M., Hern{\'a}ndez-Garc{\'\i}a, E., L{\'o}pez, C., and
Feudel, U.:
  Biological activity in the wake of an island close to a costal upwelling,
  submitted to Ecological Complexity, 2007{\natexlab{b}}.

\bibitem[{Steele and Henderson(1981)}]{Steele1981}
Steele, J. and Henderson, E.: A simple plankton model, The
American Naturalist,
  117, 676--691, 1981.

\bibitem[{Steele and Henderson(1992)}]{Steele1992}
Steele, J. and Henderson, E.: The role of predation in plankton
models, J.
  Plankton Res., 14, 157--172, 1992.


\end{thebibliography}

\end{document}